\documentclass[journal]{IEEEtran}
\usepackage{siunitx}
\ifCLASSINFOpdf
\else
\fi
\usepackage{array}
\usepackage{balance}
\usepackage{graphicx}
\usepackage{caption}
\usepackage{subcaption}
\usepackage{wrapfig}
\usepackage{amsmath}
\usepackage{epstopdf}
\usepackage{amssymb}
\usepackage{multirow}
\usepackage{cite}
\usepackage{multicol}
\usepackage{color}
\usepackage[table]{xcolor}
\usepackage{comment}

\usepackage{longtable}
\usepackage[font=footnotesize,labelfont=bf]{caption}
\newcolumntype{P}[1]{>{\centering\arraybackslash}p{#1}}
\usepackage{url}
\usepackage{hyperref}

\usepackage{array}

\usepackage{url}
\usepackage{siunitx}
\usepackage{enumerate}
\usepackage[normalem]{ulem}
\usepackage{epstopdf}

\usepackage{amssymb}% http://ctan.org/pkg/amssymb
\usepackage{pifont}% http://ctan.org/pkg/pifont
\usepackage{cancel}

\newcommand\figref{Fig.~\ref}
\begin{document}
%\title{Field Effect Transistors Based on Synthetic Two-dimensional  MoSi$_2$N$_4$}

\title{Two-dimensional~MoSi$_2$N$_4$: An Excellent 2D Semiconductor for Field-Effect Transistors}

%Investigating the p-FETs based on High Air-stable ML- Pentagonal-$\rm PdSe_2$
%Investigating the Performance of p-FETs based on High Air-stable Pentagonal-$\rm PdSe_2$
%from First-Principles Simulations
\markboth{}%
{}
\author{Keshari~Nandan,~\IEEEmembership{Graduate Student Member,~IEEE,} Barun Ghosh, Amit Agarwal, Somnath Bhowmick, and Yogesh~S.~Chauhan,~\IEEEmembership{Fellow,~IEEE}
	\thanks{This work was supported in part by the Humboldt Foundation and in part by the Swarnajayanti Fellowship (Grant No. – DST/SJF/ETA-02/2017-18) of the Department of Science and Technology~(DST), Government of India.}
	%\thanks{This work was partially funded by Swarnajayanti Fellowship and FIST Scheme of Department of Science and Technology~(DST), Government of India.}
	\thanks{K.~Nandan and Y.~S.~Chauhan are with the NanoLab, Department of Electrical Engineering, Indian Institute of Technology Kanpur,
		India, 208016, e-mail: keshari@iitk.ac.in, chauhan@iitk.ac.in.}
	\thanks{B.~Ghosh and A.~Agarwal are with the Department of Physics, Indian Institute of Technology Kanpur, India.}
	\thanks{S.~Bhowmick is with the Department of Material Science and Engineering, Indian Institute of Technology Kanpur, India.}
	%\thanks{Keshari would like to thank Barun Ghosh and Prof. Amit Verma for fruitful discussions on the characterization of material from first-principles and electron transport.}
    
}
\maketitle

%%%%%%%%%%%%%%%%%%%%%%%%%% ----- Abstract ---------%%%%%%%%%%%%%%%%%%%%%%%%%%%%%%%%%%%%%%%
\begin{abstract}
We report the performance of field-effect transistors (FETs), comprised of mono-layer of recently synthesized layered two-dimensional MoSi$_2$N$_4$ as channel material, using the first principles quantum transport simulations. The devices' performance is assessed as per the International Roadmap for Devices and Systems (IRDS) 2020 roadmap for the year 2034 and compared to advanced silicon-based FETs, carbon nanotube-based FETs, and other promising two-dimensional materials based FETs. Finally, we estimate the figure of merits of a combinational and a sequential logic circuit based on our double gate devices and benchmark against promising alternative logic technologies. The performance of our devices and circuits based on them are encouraging, and competitive to other logic alternatives.

\end{abstract}


%%%%%%%%%%%%%%%%%%%%%%%----- Section -I Intro----%%%%%%%%%%%%%%%%%%%%%%%%%%%%%%%
\section{Introduction}
\IEEEPARstart{T}{he} two-dimensional (2D) semiconductors are promising channel materials for future technology nodes, owing to the ultrathin thickness ($< \SI{1}{\nano \meter}$), no surface dangling bonds, and sharp turn-on of the density of states (DOS) at band edges\cite{Carbon,PNAS,Nobel,Chhowalla2016ER}.  In the last five years, there has been remarkable progress on synthesizing novel 2D semiconductors and several promising device concepts\cite{E-FET,DS-FET,DT-FET} have been demonstrated based on them. Also, the shortest MoS$_2$ transistor with $\SI{1}{\nano \meter}$-gate length has been fabricated, which shows a near-ideal sub-threshold swing of $\sim$ $\SI{65}{\milli \volt \per decade}$ and an ON/OFF current ratio of $\sim$ $10^6$\cite{Desai99}.
  
  %In the last five years, there has been remarkable progress in synthesizing novel 2D semiconductors and promising device concepts\cite{E-FET,DS-FET,DT-FET}, enabled by their electronic and structural properties. Also, the shortest $\SI{1}{\nano \meter}$-gate length MoS$_2$ transistor has been fabricated. This device shows a near-ideal subthreshold swing of $\sim$ $\SI{65}{\milli \volt \per decade}$ and an On/Off current ratio of $\sim$ $10^6$\cite{Desai99}.

The 2D semiconductors materials library has been enriched due to their bulk, arranged in layered form,  in which intra-layers bonds are strong covalent bonds and weak Van der Walls (vdW) force connects inter-layers. Advanced experimental techniques have been used to isolate their layers from bulk. But, most bulk materials are non-layered (strong covalent bonds connect all three dimensions). Thus, the exfoliation process can not create their 2D structure. Recently, Silicon has been introduced as a passivator during the CVD growth of non-layered molybdenum nitride ($\rm MoN_2$)\cite{Hong670}. This process results in the growth of the centimeter-scale film of layered 2D MoSi$_2$N$_4$.
% Using density functional calculations (DFT), the lowest energy structure has been investigated, and it is revealed that the system is a vdW layered 2D material formed as a stoichiometric compound with the formula MoSi$_2$N$_4$.

\begin{figure*}[t!]
	\centering
	\includegraphics[scale=0.45]{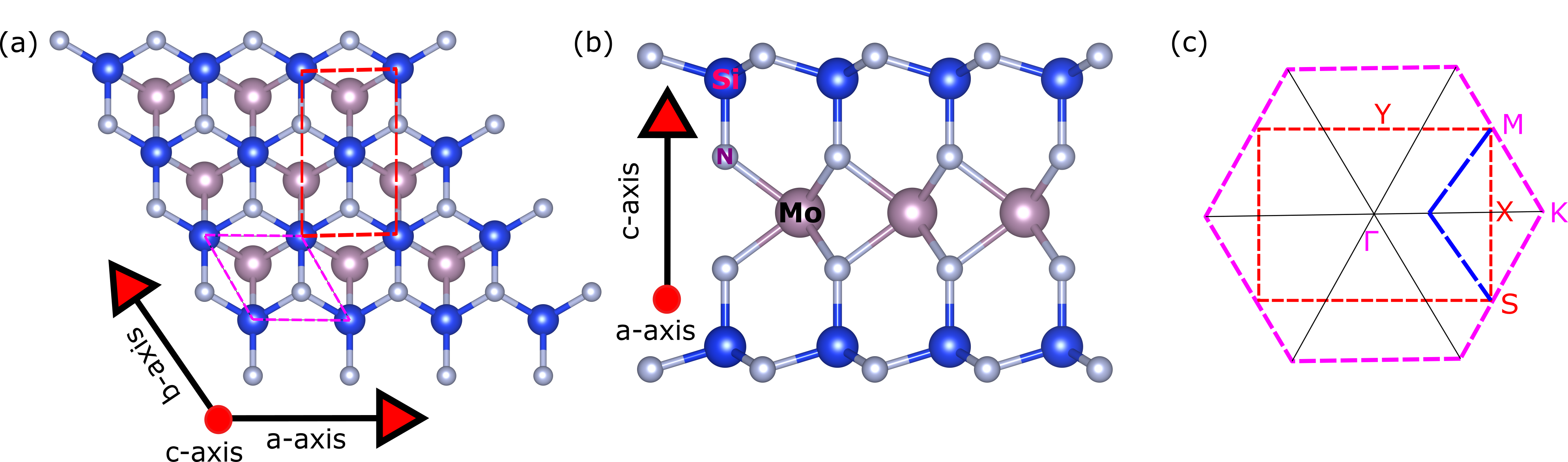}	
	\caption{(a) Top view and (b) side view of ML-MoSi$_2$N$_4$. The hexagonal primitive cell and orthogonal supercell of ML-MoSi$_2$N$_4$ are shown by dashed magenta and red lines in (a), respectively. (c) The first Brillouin zone (FBZ) and high symmetry points associated with hexagonal primitive cell (in magenta color) and orthogonal cell (in red color). The blue dashed lines in (c) show the folding of K point associated with FBZ of hexagonal cell to FBZ of orthogonal cell.}
	\label{CS}
\end{figure*}

Mono-layer (ML) MoSi$_2$N$_4$ has excellent mechanical, electronic, optical, and thermal properties\cite{Hong670,Bafekry_2021,Yu_2021,MORTAZAVI2011057126,islam2021tunable}.  It is also a promising photocatalyst for water splitting and $\rm CO_2$ reduction\cite{Bafekry_2021,MORTAZAVI2011057126}. It has been shown that ML-MoSi$_2$N$_4$ has excellent stability (far better than other 2D semiconductors), using phonon, molecular dynamics (MD) calculations, and experimental testing. Even it can be handled, processed, and stored without any protective environment, unlike black phosphorus (In ambient conditions, black phosphorus (BP) can be easily etched due to chemical degradation\cite{https://doi.org/10.1002/admi.201600121}) and MoS$_2$ (In moist air below $\SI{373}{\kelvin}$, its surface starts oxidizing\cite{doi:10.1021/j150531a020}). The Young's modulus and breaking strength for ML-MoSi$_2$N$_4$ are $\sim \SI{479}{\giga \pascal}$ and $\sim \SI{49}{\giga \pascal}$, respectively. These values are more than double of ML-MoS$_2$. ML-MoSi$_2$N$_4$ is an indirect band gap semiconductor with the experimental band gap value of $\SI{1.94}{\electronvolt}$. Its elastic constant is $\sim 4$ times of ML-MoS$_2$, and the carrier mobilities in it are $\sim 4$ times and $\sim 4$-$6$  times of ML-MoS$_2$ (the most widely studied 2D material for FETs application\cite{Radisavljevic2011,doi:10.1021/acs.nanolett.6b03999,Desai99,doi:10.1021/nl503586v,doi:10.1021/acsami.9b18577,Deblina1,Deblina2,Deblina3,PhysRevB.85.161403,doi:10.1021/nl2018178,Rastogi2014,7814267,8423445,9184047}).
Its lattice thermal conductivities\cite{MORTAZAVI2011057126,Yu_2021} are approximately 1.6 times silicon (Si) and much higher than other widely known 2D semiconductors\cite{C7NR00838D,C4CP04858J,doi:10.1063/1.4850995} (ML-MoS$_2$, As, Sb, silicene and, ML-BP), but much lower than graphene\cite{PhysRevB.80.033406,C7NR00838D}. The high lattice thermal conductivity of ML-MoSi$_2$N$_4$ ensures a high rate of heat removal through nano-electronic devices comprised of this material.
Its optical transmittance is high ($\sim 97~\% $) and comparable to graphene.
The metal contacts to the ML of MoSi$_2$N$_4$ show exceptional physical properties with a large Schottky barrier height slope parameter, outperforming most other 2D semiconductors. CMOS compatible metals (Sc and Ti) also show excellent ohmic contact to ML-MoSi$_2$N$_4$ with zero interfacial tunneling barrier\cite{Wang2021,Wu_APL}.
 Overall, Mono-layer MoSi$_2$N$_4$ is an excellent semiconductor for logic applications. However, there is a need to investigate the transport properties of FETs based on ML-MoSi$_2$N$_4$, and performance of integrated circuits (ICs) comprised of MoSi$_2$N$_4$ FETs. 

Here, we exploit the capabilities of maximally localized Wannier functions (MLWFs) \cite{MLWF1,MLWF2,MLWF3} to model electronic structure of ML-MoSi$_2$N$_4$ and generate tight-binding (TB) like Hamiltonian for the targeted device dimensions. Next, we compute transport properties of n- and p-type devices based on this ML by solving coupled Poisson and Schr$\rm \ddot{o}$dinger equations in non-equilibrium Green’s functions (NEGF) formalism. We assess the performance of MoSi$_2$N$_4$ based FETs as per the requirements of International Roadmap for Devices and Systems (IRDS) 2020 for the year 2034\cite{IRDS}. The channel length scalability of devices is also studied with their switching performance. Finally, the figure of merits (FOMs) of a combinational circuit (32-bit adder) and a sequential circuit (ALU) are estimated and benchmarked against promising logic technologies (CMOS and beyond-CMOS).

\section{Computational Methods}
Vienna Ab initio Simulation Package (VASP)\cite{VASP}, a tool based on DFT, is used to relax the atomic positions and calculate the electronic structure of ML-MoSi$_2$N$_4$. The Projector Augmented Wave (PAW)\cite{PAW} pseudopotentials with plane-wave basis set are used for DFT calculations. The generalized gradient approximation (GGA) developed by Perdew-Burke-Ernzerhof (PBE)\cite{pbe} is used to consider exchange and correlation effects. The energy cutoff of 400 eV is used for the plane-wave basis set, and the Brillouin zone integrations are performed using $12 \times 12 \times 1$ k-mesh.

The obtained Bloch/plane-wave states from DFT calculations are mapped to MLWFs using the Wannier90 suite of codes\cite{W90}. The obtained TB-like Hamiltonian in MLWF basis is used to construct the Hamiltonian for targeted device dimensions. In the transverse direction (channel width direction), periodic boundary condition (PBC) is considered with 30 uniform wave-vector samples. The constructed device Hamiltonian is used as input to solve coupled Schr$\rm \ddot{o}$dinger and Poisson equations in Non-equilibrium Green's functions (NEGF) formalism\cite{datta2005quantum,Rastogi,Tapas,9585026}. Additional computational details are described in Appendix.
 
\begin{figure}
	\centering
	\includegraphics[width=0.45\textwidth]{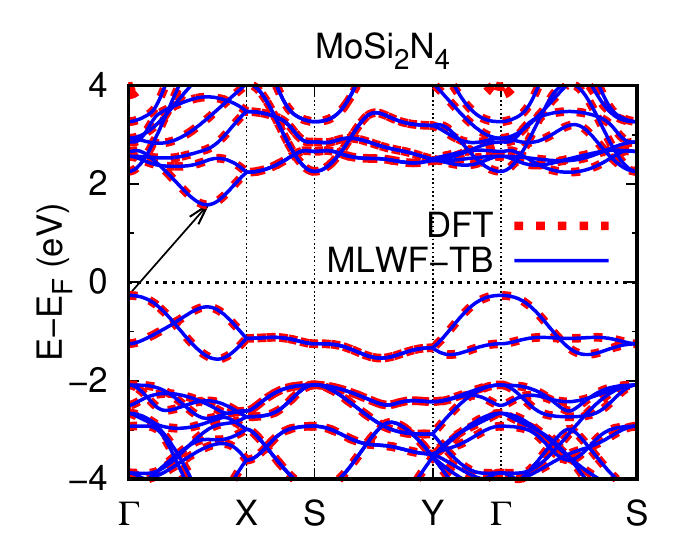}	
	\caption{Electronic band structure of MoSi$_2$N$_4$ calculated using DFT and MLWF-TB along the high symmetry path.
		% $\Gamma$-$\rm X$-$\rm S$-$\rm Y$-$\Gamma$-$\rm S$.
	}
	\label{BS}
\end{figure}
%%%%%%%%%%%%%%%%%%%%%%%--------Section --III Results---%%%%%%%%%%%%%%%%%%%%%%%%%%%%%
\section{Results}
\subsection{Structural and  Electrical Properties of ML-MoSi$_2$N$_4$ }
The two dimensional periodic replication of one Mo, two Si, and four N atoms, packed in honey-comb lattice, generate the ML of MoSi$_2$N$_4$ (see \figref{CS} (a) and (b)). This ML can be viewed as $\rm MoN_2$ ($\rm 2H-MoS_2$ like structure) sandwitched between two buckled honeycomb SiN layers (see \figref{CS} (b)). The optimized lattic constant is $a(=b) \sim$ 2.90 \AA~and the thickness of ML is $\sim$ 7.01 \AA. The optimized structural parameters agree well with the literature\cite{Li,Hong,Bafekry_2021}.

The electronic band structure of MoSi$_2$N$_4$ is plotted along with the high symmetry points ($\Gamma$-$\rm X$-$\rm S$-$\rm Y$-$\Gamma$-$\rm S$) in the orthogonal Brillouin zone (BZ). Figure. \ref{CS} (c) shows the BZ associated with hexagonal cell and orthogonal cell. Figure. \ref{BS} shows the band structure of MoSi$_2$N$_4$ obtained from DFT and MLWF-TB Hamiltonian. The band structure from MLWF-TB Hamiltonian shows a good match with DFT near VBM and CBM. It is an indirect band gap semiconductor with the conduction band maxima (CBM) lies in the way from $\Gamma$ to $\rm X$ (equivalent to $\rm K$ in hexagonal BZ) and valance band maxima (VBM) at $\Gamma$. The curvature of CBM and VBM are isotropic, but the curvature of CBM is larger than the curvature of VBM. Hence, larger effective mass for holes ($m^*_h$) than electrons ($m^*_e$) i.e., $m^*_h > m^*_e$ (see Table \ref{table1}).

\begin{figure}
	\centering
	\includegraphics[width=0.4\textwidth]{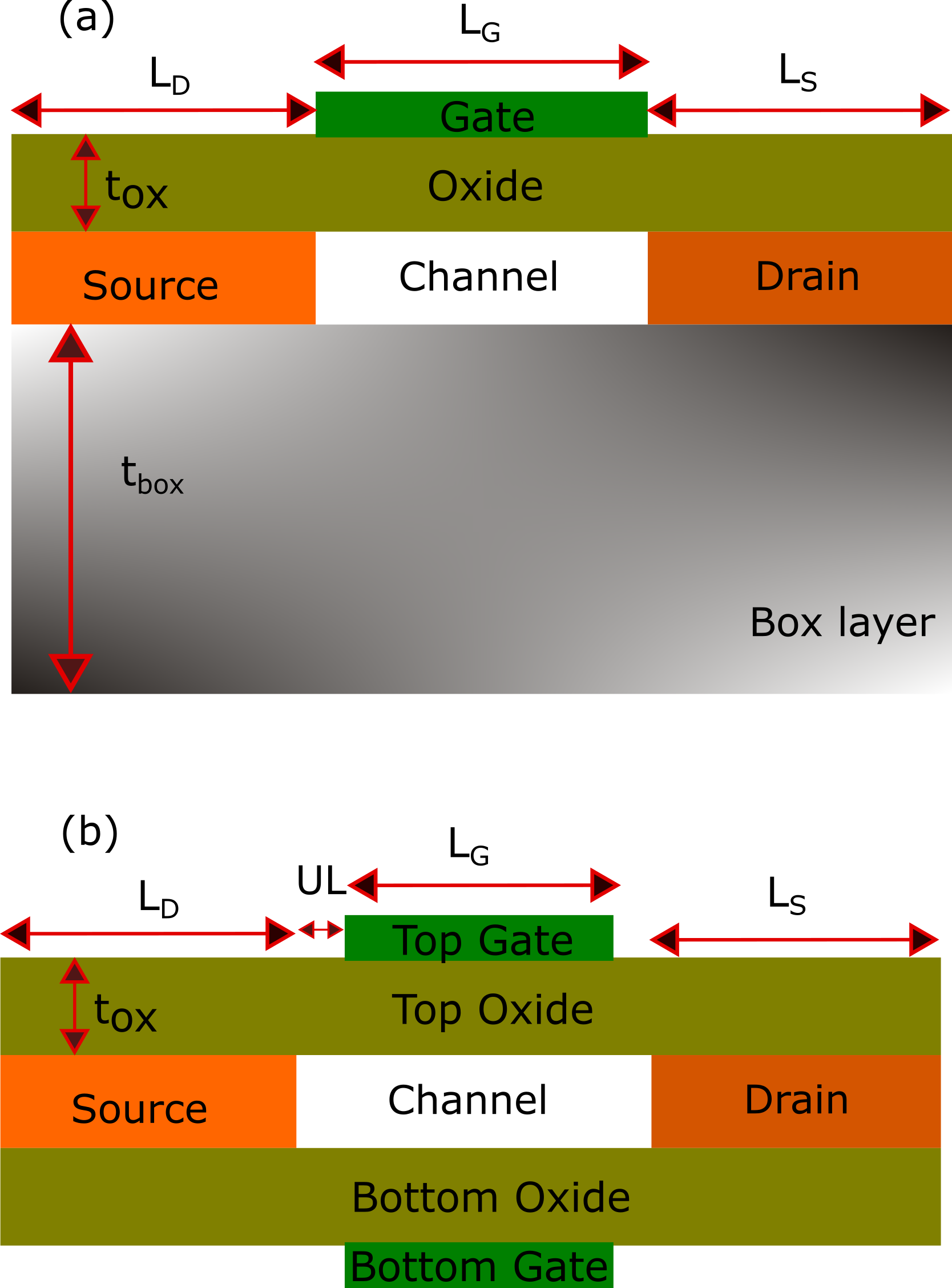}	
	\caption{Schematic of (a) single gate (SG) and (b) double gate (DG) FETs. BOX layer and gate oxides are SiO$_2$, of thickness $t_{box}$ and $\rm t_{ox}$, respectively. $L_{G}$, UL, and $L_{Ch}$~(=$~L_G + 2 \times UL$) are gate length, underlap length, and channel length,  respectively. $L_{S/D}$ is source/drain extension length. The source and drain are heavily doped with donors/acceptors for n-/p-type devices. This doping results in the energy degeneracy of 50 meV.}
	\label{DS}
\end{figure}
\subsection{Device Structure and Electrical Characteristics}
The Single gate (SG) and double-gate (DG) devices, using ML of MoSi$_2$N$_4$ as channel material, are investigated. Figure. \ref{DS} (a) and \ref{DS} (b) show the schematic of SG and DG devices, respectively. The geometrical parameters of the devices are gate length ($L_G$), channel length ($L_{Ch}$), source/drain extension $L_{S/D}$, underlap length (UL), and oxide thickness ($t_{ox}$).  For SG, $t_{box}$ is BOX layer thickness with $t_{box}$ = \SI{10}{\nano \meter}. Silicon dioxide (SiO$_2$) is used as gate oxide and BOX layer. This work aims to assess the intrinsic performance of FETs based on the mono-layer of MoSi$_2$N$_4$. Hence, the S/channel/D doping used in the device simulations are n+/undoped (intrinsic)/n+ and p+/undoped (intrinsic)/p+ for n- and p-type devices, respectively. The S/D is doped such that the Fermi level is \SI{50}{\milli \electronvolt} above/below the conduction/valance band for n-FET/p-FET. The metal gate work function is tuned to get $I_{ds}$ $\sim$ $\SI{e-2}{\mu \ampere \per \mu \meter}$ (say it OFF-current ($I_{OFF}$) at $V_{gs} = \SI{0}{\volt}$ for all the simulated devices. 
%\textcolor{blue}{For $I_{ON}$, the drain current is estimated at $V_{gs}$ = $V_{gs}^{ON} (=V_{gs}^{OFF}+V_{DD})$. For the other $I_{OFF}$ values, $V_{gs}^{OFF}$ is estimated such that at $V_{gs}$ = $V_{gs}^{OFF}$, $I_{ds}$ = $I_{OFF}$, and further $I_{ON}$ is estimated.}

%(\textcolor{blue}{say it $V_{gs}^{OFF}$ for $I_{OFF}$ $\sim$ $\SI{e-2}{\mu \ampere \per \mu \meter}$})

\begin{figure}[t]
	\includegraphics[width=0.5\textwidth]{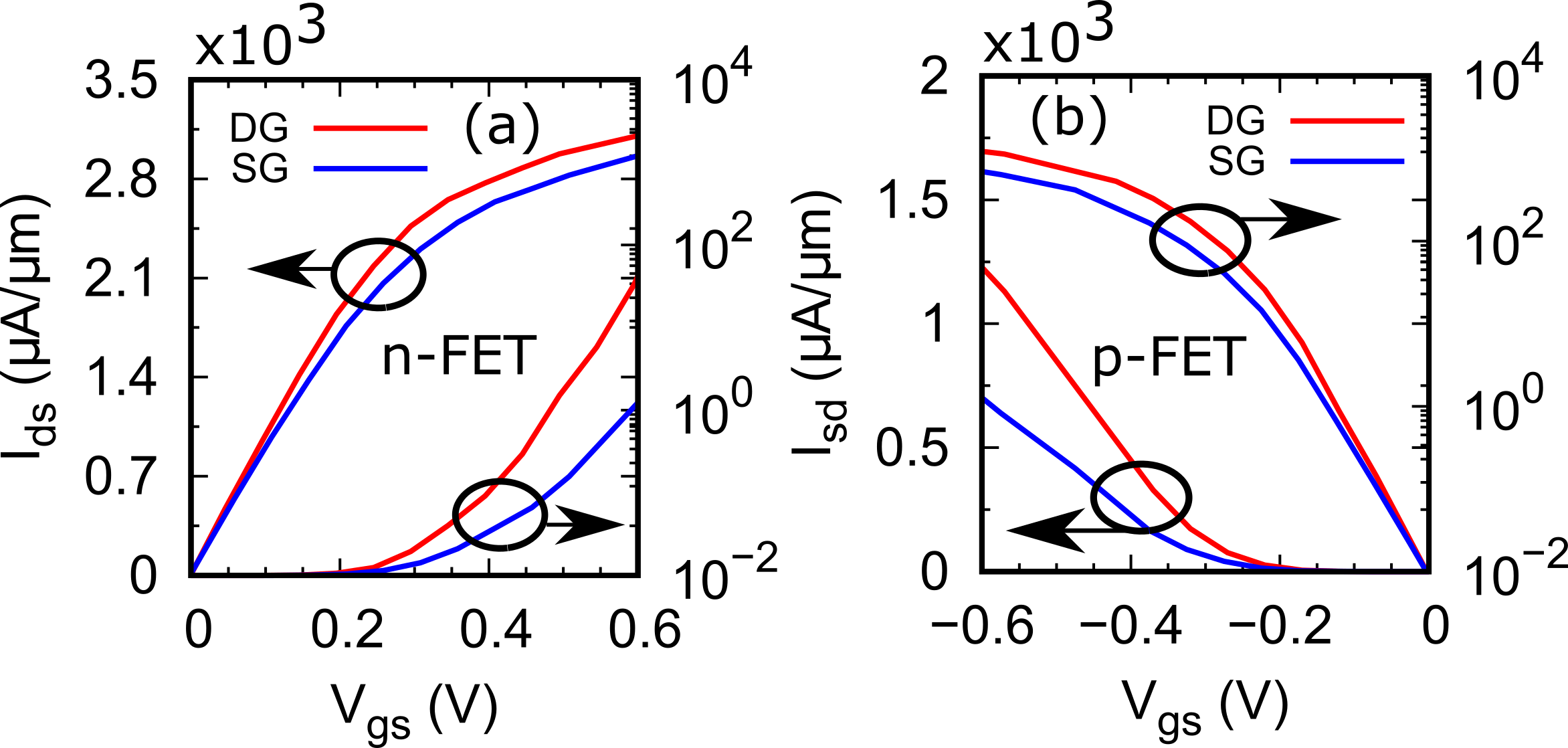}	
	\caption{Transfer characteristics of (a) n-FET and (b) p-FET using MoSi$_2$N$_4$ as channel material. $\rm L_{Ch}(=\rm L_{G}) = 12~nm$, $\rm t_{ox} = 0.6~nm$, and $\rm V_{DD} = 0.50~V$. The $\rm I_{OFF}$ is fixed at $10^{-2} \mu A / \mu m$ for all devices.}
	\label{IV_12nm}
\end{figure}

We start the investigation by simulating SG and DG devices with $L_{Ch}(=L_{G}) = \SI{12}{\nano \meter}$, equivalent oxide thickness (EOT) = $\SI{0.60}{\nano \meter}$,~and~$V_{DD} = \SI{0.50}{\volt}$. The source-to-drain tunneling is negligible for the devices with $L_{Ch} = \SI{12}{\nano \meter}$, and the band-to-band tunneling is negligible due to the high value of band gap in this ML. These device parameters are considered according to IRDS 2020 roadmap for the year 2034. According to the roadmap, the expected channel materials are germanium (Ge) and 2D materials, and the expected devices are 2D materials based devices and FeFETs for the year 2034\cite{IRDS}. Figure. \ref{IV_12nm} shows the transfer characteristics for n-FETs and p-FETs. In both types, DG configuration shows a steeper sub-threshold slope (SS) and higher $I_{ON}$ than SG. 

 The SS for thermionic current is proportional to $1+C_q/C_{OX}$\cite{lundstrom2006nanoscale}, where $C_q$ is quantum capacitance and $C_{OX}$ is oxide capacitance. For DG configuration, oxide capacitance is double than SG, hence DG devices show better SS than SG for $L_{Ch} = \SI{12}{\nano \meter}$. The stepper SS and higher value of $C_{OX}$ in DG ensure higher $d|I_{ds}|/d|V_{gs}|$ and higher mobile carrier concentration, respectively than SG, hence higher $I_{ON}$ than SG. Defects and series parasitics resistance from contacts and S/D access region are not included in our simulations, usually limiting the device's performance.

%Figure. \ref{DS} shows the diagrammatic representation of the studied devices. The devices are single gate (SG) and double gate (DG). The geometrical parameters of the devices are gate length ($\rm L_G$), channel length($L_{Ch}$), source/drain extension $\rm L_{S/D}$, underlap length (UL), and oxide thickness ($\rm t_{ox}$). For SG, $\rm t_{box}$ is BOX layer thickness. $\rm SiO_2$ is used as gate oxide and BOX layer. The channel thickness ($h$) is $\sim$ 0.7 nm and it is composed of synthetic two-dimensional material MoSi$_2$N$_4$.
%%%%%%%%%%%%%%%%%%%%%%%%%----- Section - IV Con -----%%%%%%%%%%%%%%%%%%%%%%%%%%%%

\subsubsection{Channel length Scaling}
\begin{figure}[t]
 	\centering
 	\includegraphics[width=0.45\textwidth]{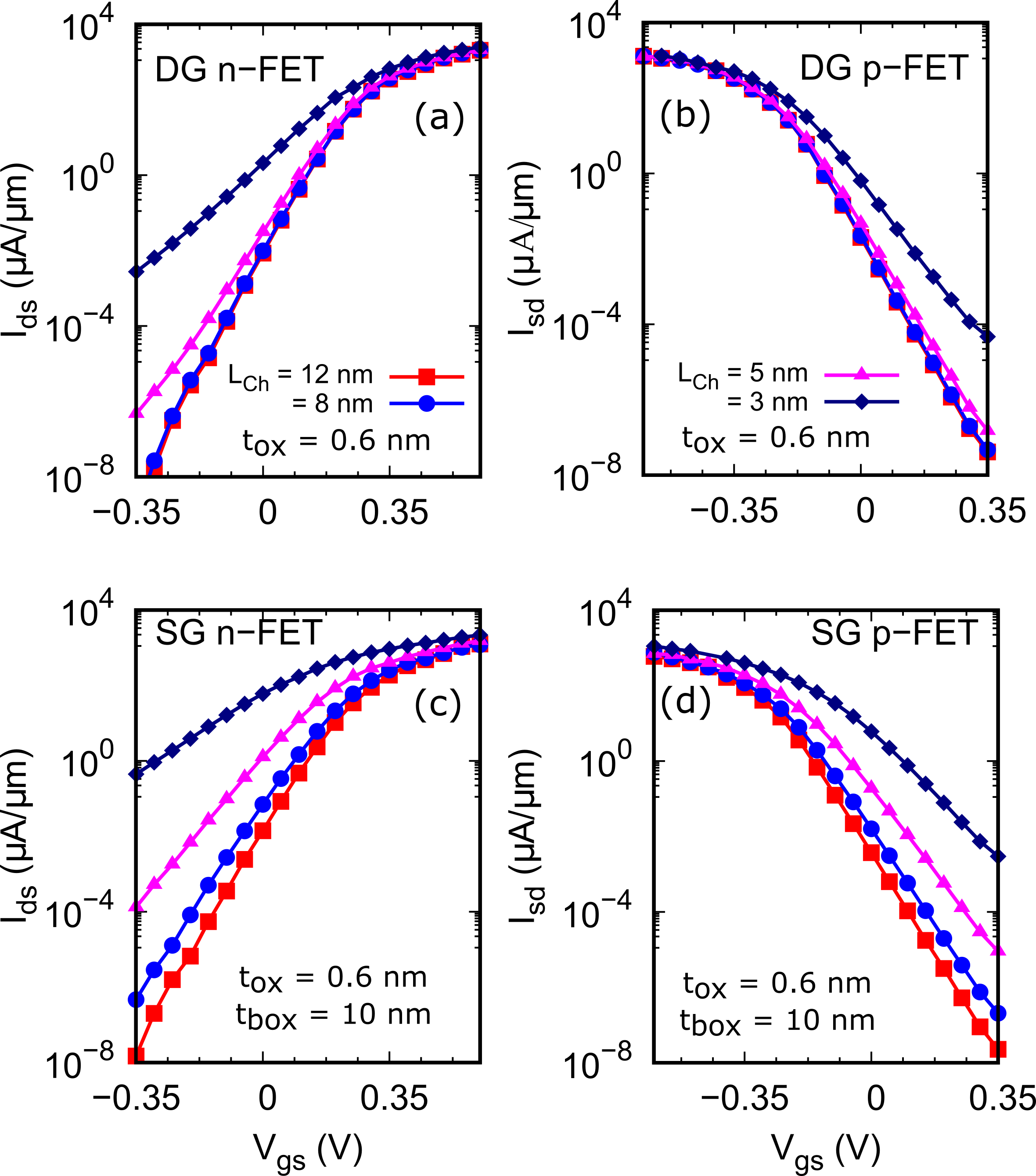}	
 	\caption{Transfer characteristics of double gate: (a) n-FET (b) p-FET and single gate: (c) n-FET (d) p-FET for different channel lengths.}
 	\label{LCH_Scaling}
 \end{figure}

The channel length scaling is performed to study the scalability and immunity to source-to-drain tunneling (SDT) of ML-MoSi$_2$N$_4$ based FETs. The SDT plays a significant role in deteriorating device performance at short channel lengths. It is more severe for the low effective mass carriers than high effective mass, as the tunneling probability\cite{griffiths2018introduction} is proportional to $\exp  (-\sqrt{m^*})$. Figures. \ref{LCH_Scaling} (a), (b), (c), and (d) show the transfer characteristics of DG n-FET, DG p-FET, SG n-FET, and SG p-FET, respectively for various channel lengths $L_{Ch}$ = 12, 8, 5, 3 nm.

\begin{table}[!b]
	\renewcommand{\arraystretch}{1.2}
	\caption{Lattice parameters ($a(=b)$), Electronic Band Gap ($\rm E_G$), electron/hole effective mass ($m^*_{e}/m^*_{h}$) and location of CBM/VBM for ML of MoSi$_2$N$_4$ }
	\label{tab:example}
	\centering
	\begin{tabular}{c|c|c|c|c|c|c}
		\hline
		\rowcolor{lightgray}  &$a$ (\AA)&$h$(\AA) & $\rm E_G (eV)$ &$m^*_{e}$&$m^*_{h}$&VBM/CBM \\
		\hline
		\hline
		MoSi$_2$N$_4$ &2.90& 7.01&  1.842 & 0.478&1.184 &$\Gamma$ / $\Gamma$-X \\

		\hline
	\end{tabular}
	\label{table1}
\end{table}
%%%%%%%%%%%%%%%%%%%%%%%%%%% Acknowledgement %%%%%%%%%%%%%%%%%%%%%%%%%%%%%%%%%%%%%%%%%

The impact of SDT is more for n-type devices than p-type because the curvature of CBM is greater than VBM. For each type, DG is more immune to SDT than SG as DG shows better gate controllability than SG. Hence, DG devices are more scalable than SG.
The best scalable device is DG p-FET, and it can be scaled down to 5 nm. At $L_{Ch}$ = $\SI{5}{\nano \meter}$, DG p-FET show SS $\sim$ 65 and $I_{ON}/I_{OFF}$  $> 3 \times 10^{6}$ for LP applications; For HP applications, it shows SS $\sim$ 67 mV/decade and $I_{ON}/I_{OFF}$   $> 10^4$.

%the scalability of DG p-FET is better than others and can be scalable down to 5 nm. At $L_{Ch}$ = $\SI{5}{\nano \meter}$, DG p-FET show SS $\sim$ 65 and $\rm I_{ON}/I_{OFF}$  $> 3 \times 10^{-6}$ for LP applications; For HP applications, it shows SS $\sim$ 67 mV/decade and $\rm I_{ON}/I_{OFF}$   $> 10^4$. 

\begin{figure}[!b]	
	\centering
	\includegraphics[width=0.3\textwidth]{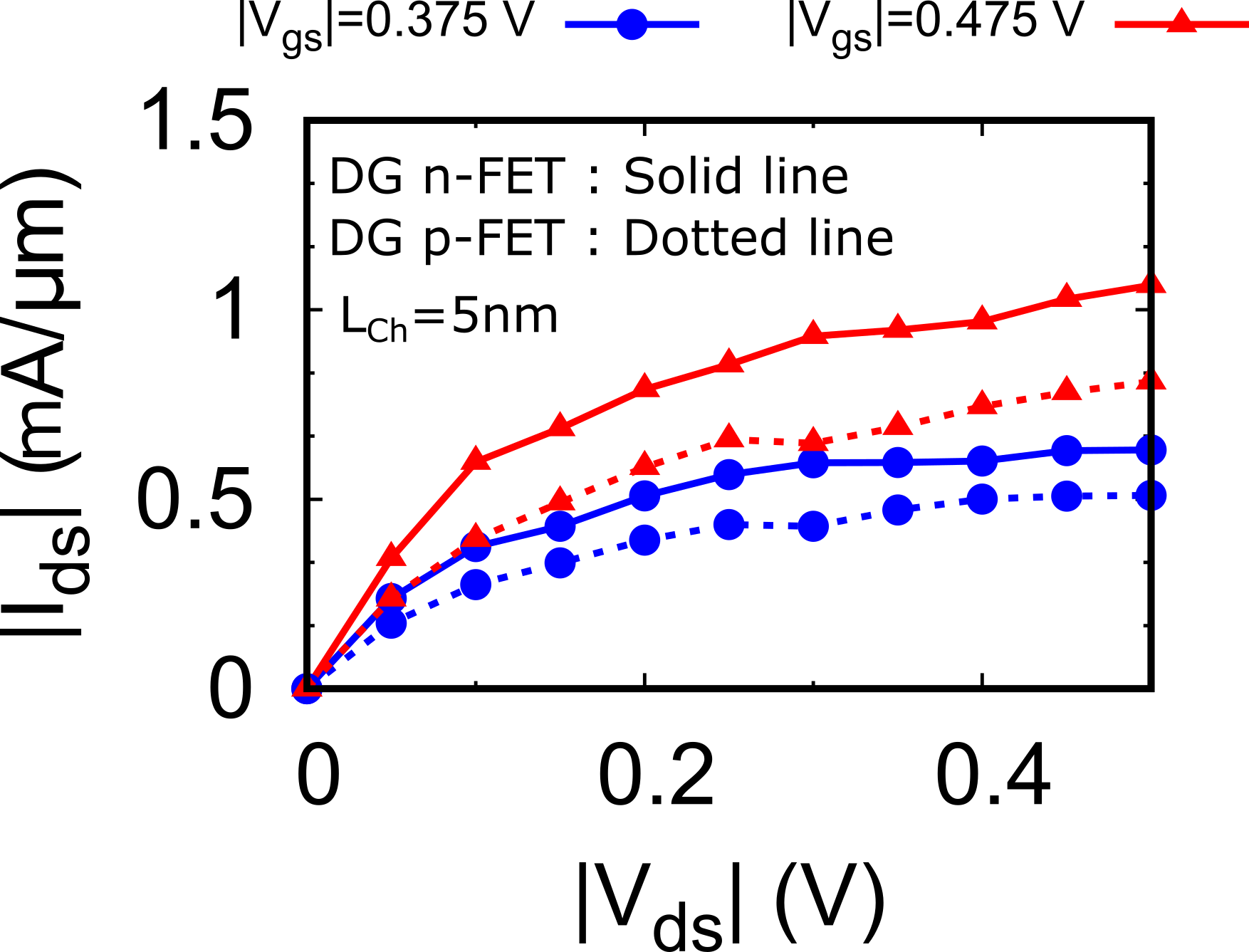}	
	\caption{Output characteristics of DG n-FET and p-FET for $L_{Ch}(=L_G) = 5~nm$.}
	
	\label{out}
\end{figure}

\begin{figure}[t]
 	\centering
 	\includegraphics[width=0.45\textwidth]{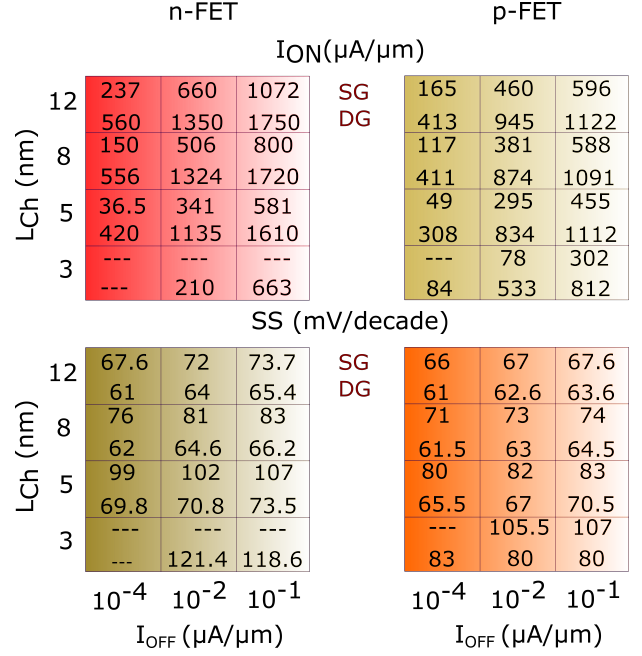}	
 	\caption{ $I_{ON}$ and sub-threshold swing (SS) of n-FET and p-FET vs $L_{Ch}$ for three different values of $I_{OFF}$.}
 	\label{fig:ION_SS}
 \end{figure}

\begin{figure}[!b]	
 	\centering
 	\includegraphics[width=0.45\textwidth]{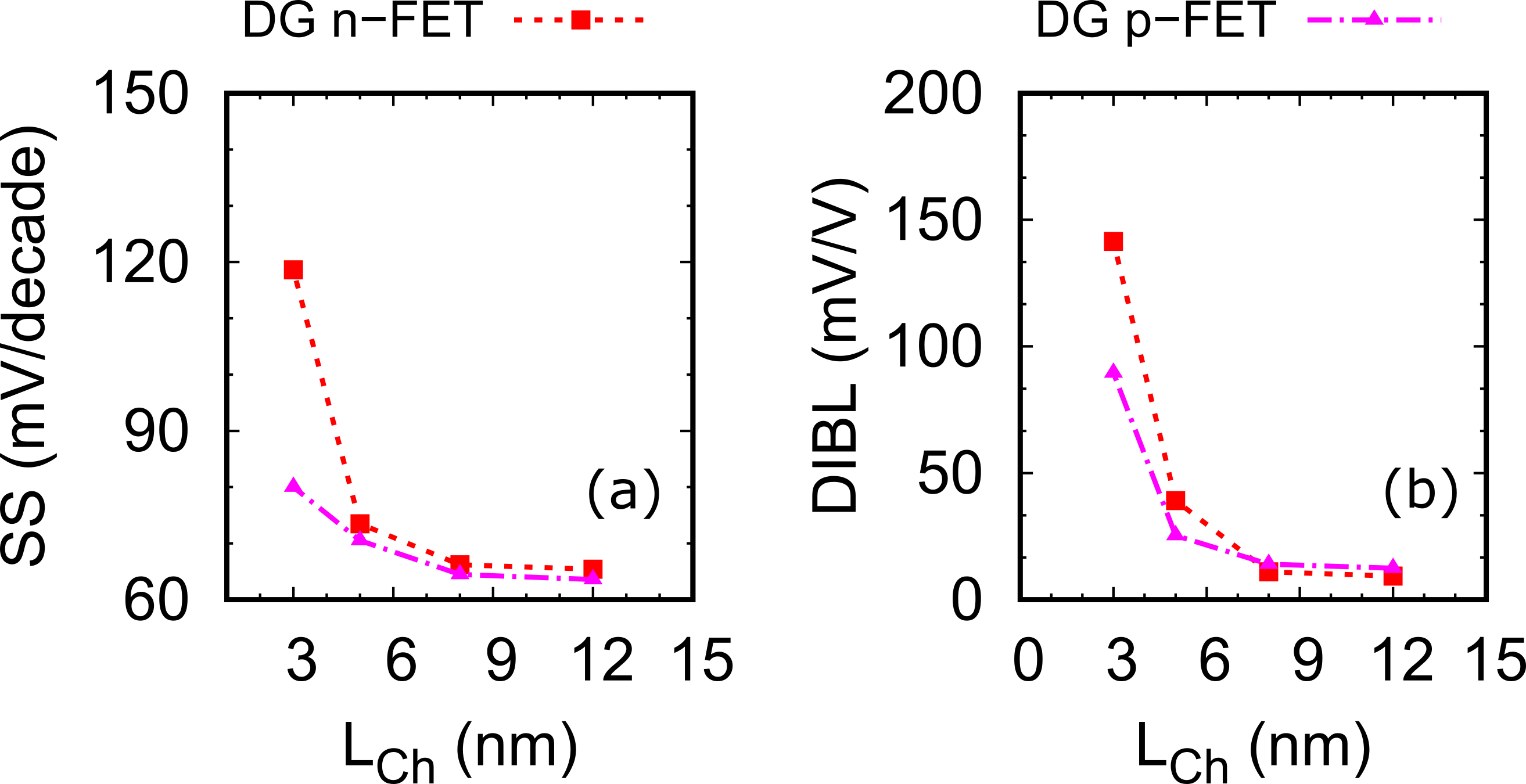}	
 	\caption{Variation in (a) SS and (b) DIBL for double-gate devices with channel length.} 
 	
 	\label{SS_DIBL}
 \end{figure}

\begin{figure}[!t]
 	\centering	
 	\includegraphics[width=0.45\textwidth]{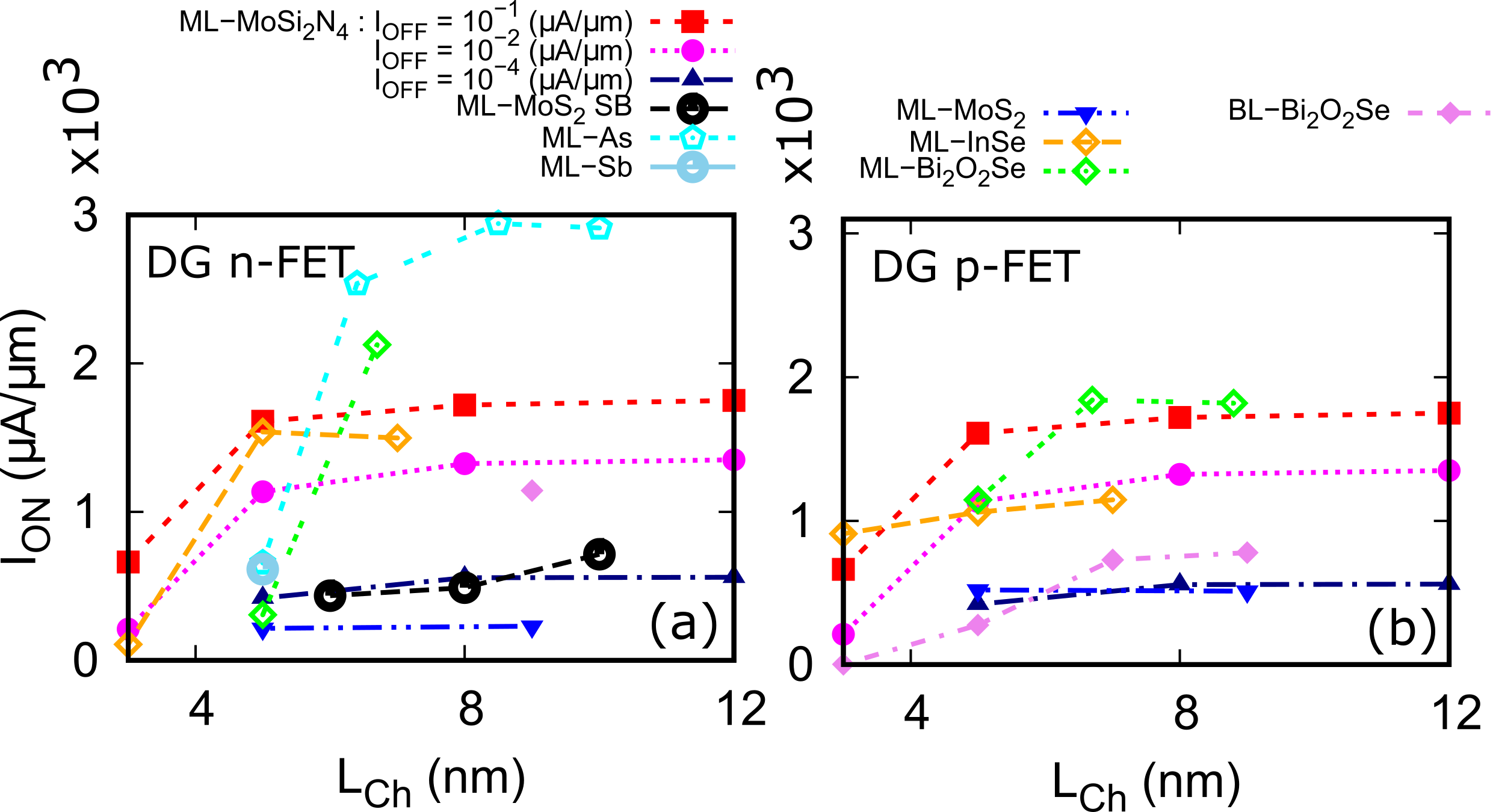}	
 	\caption{ ON current vs $L_{Ch}$ for DG (a) n-FET and (b) p-FET devices. The data for several promising 2D materials\cite{ML_Bi2O2Se,InSe,BL_Bi2O2Se,As_Sb} based DG devices, obtained from first-principle calculations ($I_{OFF} = \SI{e-1}{\mu \ampere \per \mu \meter}$), are also shown in same plot for comparison.} 
 	
 	\label{ION_LCH}
 \end{figure}

Figure. \ref{out} shows the output characteristics of DG n-FET and p-FET with $L_{Ch}$ = 5 nm.
Figure. \ref{fig:ION_SS} shows $I_{ON}$ and SS of n-FET and p-FET for various $L_{Ch}$ at three different values of $I_{OFF}$.
Also, for a full assessment of DG devices at short channel lengths, DIBL is estimated from the change in threshold voltage ($V_{th}$) obtained by varying $V_{ds}$ from $V_{low}$ (= 50 mV) to $V_{DD}$. Figures. \ref{SS_DIBL} (a) and (b) show the variations of SS and DIBL with $L_{Ch}$, respectively.
Figure. \ref{ION_LCH} shows the variation of $I_{ON}$ vs. $L_{Ch}$ for DG n- and p-FETs based on ML-MoSi$_2$N$_4$ and other promising 2D materials.
 For both types, the ML-BP FETs show the highest $I_{ON}$ \cite{Yin2015,BP} among others (the data is not shown in the figure). But, the stability of BP remains the primary concern\cite{https://doi.org/10.1002/admi.201600121}. For n-type devices: (1) For $L_{Ch}> \SI{5}{\nano \meter}$, ML-As, and ML-Bi$_2$O$_2$Se have high $I_{ON}$ than ML-MoSi$_2$N$_4$. However, the preparation of As can produce toxic arsenic trioxide, making its fabrication more complex, other than stability issue\cite{C7CS00125H}. ML-Bi$_2$O$_2$Se shows good environmental stability\cite{Wu2017}. But the short-channel effects are more in ML-As and ML-Bi$_2$O$_2$Se than ML-MoSi$_2$N$_4$. (2) ML-InSe shows the lower value of $I_{ON}$ than ML-MoSi$_2$N$_4$ and the  $I_{ON}$  difference increases as we go below $\SI{5}{\nano \meter}$ channel length. However, compelling surface oxidation in InSe deteriorates mobility and causes uncontrollable current hysteresis in InSe FETs\cite{doi:10.1021/acsnano.7b03531}. (3) The ML-MoSi$_2$N$_4$ shows superior ON-state current than ML-MoS$_2$ FETs. For p-type devices: (1) Except InSe, the MoSi$_2$N$_4$ shows superior ON-state current than others for $L_{Ch} < \SI{5}{\nano \meter}$. (2) The $I_{ON}$ for ML-MoSi$_2$N$_4$ is very close to ML-Bi$_2$O$_2$Se and superior to others for  $L_{Ch} > \SI{5}{\nano \meter}$. BL-Bi$_2$O$_2$Se suffers from high leakage current due to small bandgap ($\rm E_G$ $\sim$ $\SI{0.18}{\electronvolt}$), and $I_{ON}$ is much lower than ML-MoSi$_2$N$_4$ for both n- and p-type devices.
 
 
 \begin{figure}[!b]	
 	\centering
 	\includegraphics[width=0.45\textwidth]{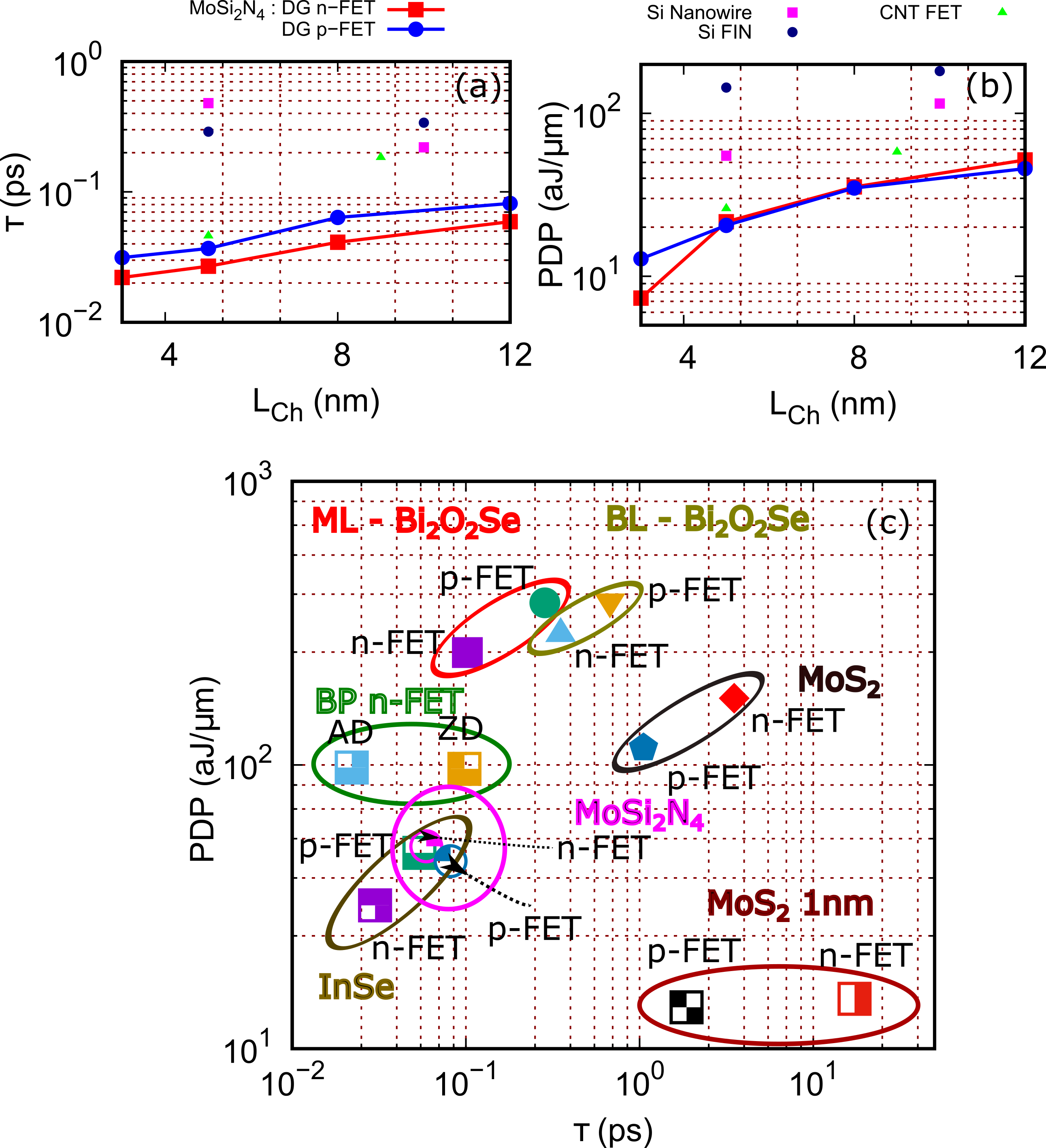}	
 	\caption{(a) $\tau$ and (b) PDP vs $L_{Ch}$ for MoSi$_2$N$_4$ FETs. (c) Benchmarking of MoSi$_2$N$_4$ FETs against promising 2D materials based FETs. The performance of 1nm-$L_G$ MoS$_2$ FET is also shown for comparison. }
 	\label{PDP}
 \end{figure}

\begin{figure}[t]
	\centering
	\includegraphics[width=0.45\textwidth]{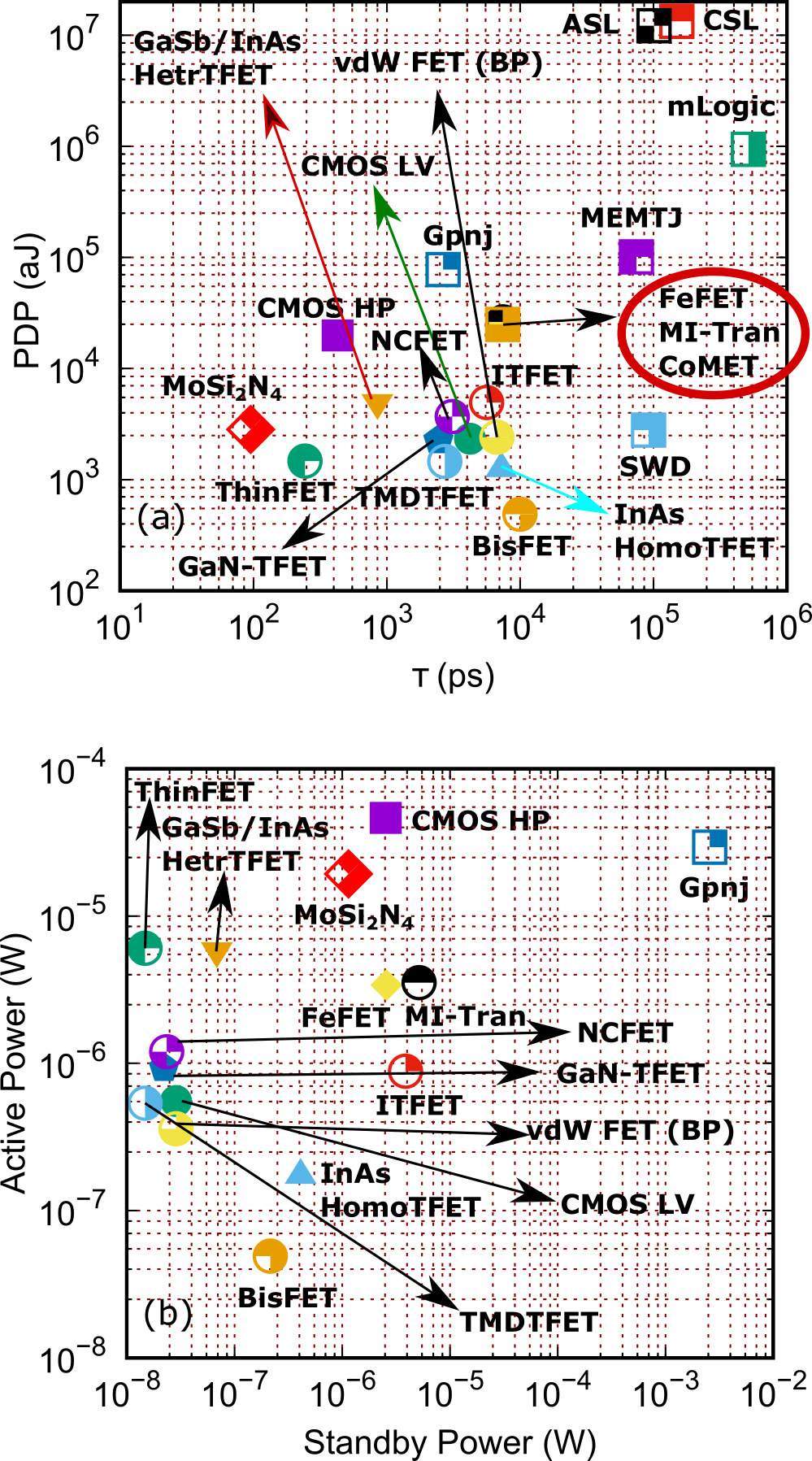}
	\caption{Comparison of (a) PDP vs $\tau$ and (b) active power vs standby power for a 32-bit adder based on ML-MoSi$_2$N$_4$ FETs against other promising logic technologies\cite{Nikonov2015}.}
	\label{32_bit_adder}
\end{figure}

\subsubsection{Switching Performance of FETs}
The delay and power dissipation product (PDP) are crucial figures of merits (FOMs) for logic applications. They determine switching speed and switching energy (energy per switching event), respectively. Figure. \ref{PDP} (a) and (b) show the delay [$\tau = (Q_{ON}-Q_{OFF})/I_{ON}$] and PDP [$=(Q_{ON}-Q_{OFF})V_{DD}$] vs $L_{Ch}$, respectively for DG FETs. These switching parameters for advance Si FETs (Si FinFETs\cite{FIN1,FIN2} and Si Nanowire FETs\cite{NW1,NW2}) and carbon nanotube (CNT) FETs\cite{CNT1,CNT2} are also shown in Fig. \ref{PDP}. For MoSi$_2$N$_4$ FETs with $L_{Ch}=3,~5,~8,~12~nm$, $\tau$ and PDP lies in range $\rm 0.022-0.082~ps$ and $\rm 7.33 - 51.7~aJ/ \mu m$, respectively. Overall, MoSi$_2$N$_4$ FETs switch faster with lower switching energy than advance Si FETs ($\rm 0.22-0.48~ps$,~$\rm 55-182~aJ/ \mu m$) and CNT FETs ($\rm 0.046-0.184~ps$,~$\rm 26-58~aJ/ \mu m$). It is worth to notice that the theoretical limit on delay for non-tunneling barrier controlled binary logic switch is $\SI{0.04}{\pico \second}$\cite{NW1}. Our non-tunneling devices ($L_{Ch} =\SI{8}{\nano \meter}$ and $\SI{12}{\nano \meter}$) have $\tau$ in the range $\rm 0.041-0.082~ps$.

% These values are better than advance Si FETs ($\rm 0.22-0.48~ps$,~$\rm 55-182~aJ/ \mu m$) and CNT FETs ($\rm 0.046-0.184~ps$,~$\rm 26-58~aJ/ \mu m$). Overall, switching performance of MoSi$_2$N$_4$ FETs are better than advance Si FETs and CNT FETs. It is worth to notice that the theoretical limit on delay for non-tunneling barrier controlled binary logic switch is $0.04~ps$\cite{NW1}. Our non-tunneling devices ($L_{Ch} =8$ and $12 ~nm$) have $\tau$ in the range $\rm 0.041-0.082~ps$.

The switching performance of DG MoSi$_2$N$_4$ based FETs ($L_{Ch}=\SI{12}{nm}$) is benchmarked against other promising 2D materials based DG FETs\cite{ML_Bi2O2Se,BL_Bi2O2Se,InSe,BP}. Other than 1 nm-MoS$_2$ FET, the channel length of other 2D materials based FETs are in the range $\rm 7-9~nm$. The five best performing devices are BP AD n-FET, InSe n-FET, InSe p-FET, MoSi$_2$N$_4$ n-FET, and MoSi$_2$N$_4$ p-FET, these have energy-delay product (EDP = PDP $\times \tau$) $<$ ($4 \times \SI{e-30}{\joule \second \per \mu \meter} $). The BP armchair direction (AD) n-FET has best switching speed among these, followed by InSe n-FET, InSe p-FET, MoSi$_2$N$_4$ n-FET, and MoSi$_2$N$_4$ p-FET. The MoS$_2$ 1nm-$L_G$ FETs have lower switching energy than others, but p-FET and n-FET are $\sim$ 23 times and $\sim$ 212 times slower than MoSi$_2$N$_4$ p-FET, respectively. The EDP of our devices are close to high performing InSe p-FET ($\SI{2.64e-30}{\joule \second \per \mu \meter}$) and BP AD n-FET ($\SI{2.15e-30} {\joule \second \per \mu m}$). The best performing device is InSe n-FET with approximately one third EDP than our devices, and EDP of BP zigzag (ZZ) n-FET is approximately three times than our devices. Others (Bi$_2$O$_2$Se and MoS$_2$) have EDP ($25-603 \times \SI{e-30}{\joule \second \per \mu \meter} $) far from MoSi$_2$N$_4$ FETs. All the devices have $I_{OFF} =$ $\SI{e-1}{\mu A \per \mu m}$.

 % The energy-delay product ($EDP=PDP \times \tau$) of DG n-FET and p-FET are $\rm 3.05 \times 10^{-30}~ (J.s)/ \mu m$ and $\rm 3.74 \times 10^{-30} ~(J.s)/ \mu m$, respectively. 

\subsubsection{FOMs of 32-bit Adder and ALU}
The FOMs of a combinational (32-bit adder) and a sequential logic circuit (32-bit ALU) based on MoSi$_2$N$_4$ FETs are estimated, using the methodology of \cite{Nikonov2015,Pan2017}. The n- and p-type DG MoSi$_2$N$_4$ FETs with $ L_{Ch} =\SI{12}{\nano \meter}$ are considered, and are sized to deliver same current. The $\SI{15}{\nano \meter}$ metal half pitch and $\SI{60}{\nano \meter}$ contacted gate pitch are taken, which results in interconnect capacitance ($C_{ic}$) =  $\SI{0.0378}{\femto \farad}$ and interconnect delay ($t_{ic}$) = $\SI{0.1891}{\pico \second}$ for $I_{OFF} = \SI{e-1}{\mu \ampere \per \mu \meter}$ (HP applications).

\begin{figure}	
	\centering
	\includegraphics[width=0.45\textwidth]{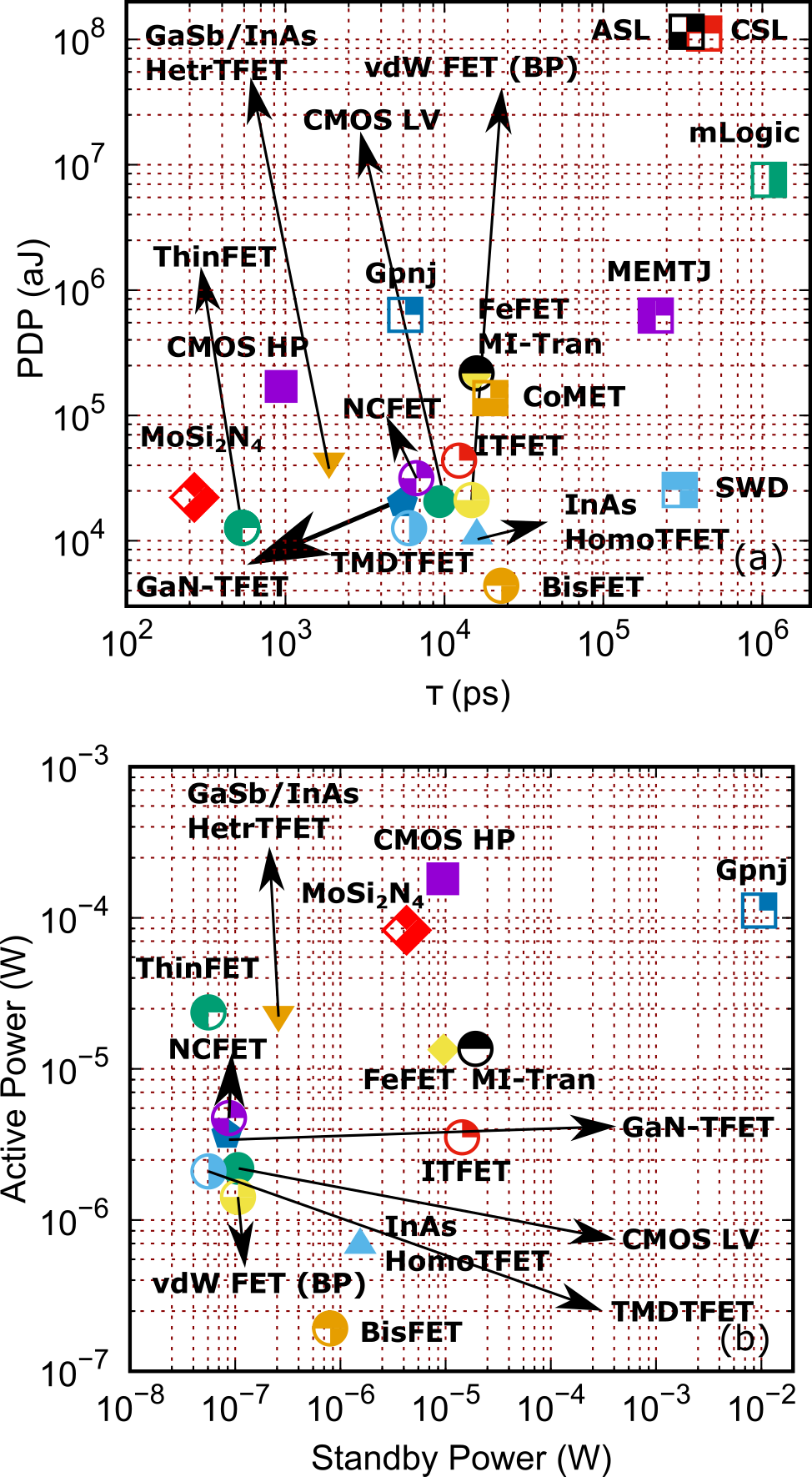}	
	\caption{Comparison of (a) PDP vs $\tau$ and (b) active power vs standby power for a 32-bit ALU based on ML-MoSi$_2$N$_4$ FETs against other promising logic technologies\cite{Nikonov2015}. The performance of ALU comprised of MoSi$_2$N$_4$ FETs is comparable to CMOS-HP.}
	\label{alu}
\end{figure}

 Using these data, the $\tau$ and $\rm PDP$ are calculated for inverter with fan-out of four ($\tau \times$ PDP = $\SI{0.3984}{\pico \second} \times \SI{21.0382}{a \joule}$), NAND gate with fan-in of four ($\SI{0.6379}{\pico \second} \times \SI{19.4914}{a \joule}$), 32-bit adder ($\SI{111.32e2}{\pico \second} \times \SI{2.1885e3}{a \joule}$), and ALU ($\SI{267.41}{\pico \second} \times \SI{2.2136e4}{a \joule}$) comprised of MoSi$_2$N$_4$ FETs. Figure \ref{32_bit_adder} (a) and \ref{alu} (a) show $\tau$ vs $\rm PDP$ for 32-bit adder and ALU, respectively based of MoSi$_2$N$_4$ and other spintronics and electronics logic devices (these data are taken from \cite{Pan2017}). The spintronics devices (ASL, CSL, MEMTJ, mLgic, CoMET, and SWD) based adder and ALU switch slower and consumes more energy than electronics devices. Among electronics devices, switching speed of MoSi$_2$N$_4$ based adder and ALU is comparable to CMOS HP and ThinFET ($\rm WTe_2/SnSe_2$ heterojunction interlayer TFET), but the switching energy of MoSi$_2$N$_4$ is one order less than CMOS HP and comparable to TFET devices (ThinFET and $\rm GaSb/InAs$ heterojunction TFET). Further, we characterize the power consumption parameters of adder and ALU. Figures \ref{32_bit_adder} (b) and \ref{alu} (b) show the standby power vs. active power for adder and ALU comprised of MoSi$_2$N$_4$ FETs and other promising electronic devices. The standby power and active power of the circuits consisting of MoSi$_2$N$_4$ devices are close to  CMOS HP and more than TFET devices (ThinFET and $\rm GaSb/InAs$ heterojunction TFET).

%($W_p/W_n \sim 1.6$) 

\section{Conclusion}

%Two-dimensional materials are promising candidates in the material engineering aspect for future technology nodes. 
Two-dimensional semiconductors are promising candidates as a channel of next-generation electronic devices. In recent years, a new 2D semiconductor with the formula MoSi$_2$N$_4$ has been discovered and gaining attention owing to its excellent physical and electronic properties. In this work, the performance of FETs based on recently discovered ML-MoSi$_2$N$_4$ is assessed, using the first principles based quantum transport simulations. The upper-performance limit is reported, as the transport is assumed to be ballistic in nature. The scalability and impact of source-to-drain tunneling are investigated by performing channel length scaling study. The double-gate devices are scalable down to $\SI{5}{\nano \meter}$. However, the p-type devices are more immune to short channel effects (SCEs) than n-type. The performance is estimated as per the IRDS roadmap for the year 2034. The key FOMs for logic switches are calculated and benchmarked against other promising 2D materials-based FETs. It is found that the switching parameters of double-gate devices are better than advanced Si-FETs and CNT-based FETs. Finally, we calculate the FOMs of a combinational and a sequential logic circuit based on our double gate devices and benchmark against CMOS and beyond-CMOS logic technologies. The performance of 32-bit adder and ALU are promising among other alternative logic technologies.

%%%%%%%%%%%%%%%%%%%%%%%%%% Appendix %%%%%%%%%%%%%%%%%%%%%%%%%%%%%%%%%%%%%%%%%%%%%%%%%%%%%
\section*{Appendix}
At each bias ($V_{gs},V_{ds}$), the transmission coefficient can be expressed as,

\begin{align}\label{Transmission}
T(E,k,V_{gs},V_{ds}) = Trace[\Gamma_S G^R \Gamma_D G^A]~.
\end{align}
Here, $G^R$ and $G^A$ are retarded and advance Green functions, respectively. $\Gamma_S/D$ is the broadening from source/drain contacts. $G^R$, $\Gamma_{L/R}$, and $G^A$ can be expressed as,

\begin{subequations}
	
	\begin{align}
	G^R(E,k,V_{gs},V_{ds}) = [EI-H(k)-\Sigma]^{-1}~,\\
	\Gamma_{S/D}(E,k,V_{gs},V_{ds}) = i[\Sigma_{S/D}-\Sigma^\dagger_{S/D}]~,\\
	G^A(E,k,V_{gs},V_{ds}) =[G^R(E,k,V_{gs},V_{ds})]^\dagger~,
	\end{align}
\end{subequations}
where, E and I are energy and identity matrix, respectively. $\Sigma$ (= $\Sigma_{S}$+$\Sigma_{D}$) is the sum of source and drain contact self-energy matrix, and H(k) is the channel Hamiltonian.

\begin{comment}

\begin{align}
\nonumber
G^R(E,k,V_{gs},V_{ds}) = [EI-H(k)-\Sigma]^{-1}
\end{align}
%%
\begin{align}
\nonumber
\Gamma_{S/D}(E,k,V_{gs},V_{ds}) = i[\Sigma_{S/D}-\Sigma^\dagger_{S/D}]
\end{align}
%%
\begin{align}
\nonumber
G^A(E,k,V_{gs},V_{ds}) =[G^R(E,k,V_{gs},V_{ds})]^\dagger 
\end{align}
\end{comment}

Next, the drain current ($I_{ds}$) is calculated using the Landauer-$\rm B\ddot{u}ttiker$ approach\cite{Landauer}. For a given gate-to-source voltage ($V_{gs}$) and drain-to-source voltage ($V_{ds}$), it can be expressed as,

\begin{align}\label{Landauer_formulation}
\nonumber
I_{ds}(V_{gs},V_{ds}) = \frac{e}{\pi \hbar} \int_{-\infty}^{\infty} \sum_{k} T(E,k,V_{gs},V_{ds})\\ [f(E-\mu_s)-f(E-\mu_d)] dE~,
\end{align}
where, $e$ is the electron charge, $\hbar$ is the reduced Planck constant, $T(E,V_{gs},V_{ds})$ = $\sum_{k} T(E,k,V_{gs},V_{ds})$ is the transmission coefficient at Energy $E$ for a given bias ($V_{gs}$, $V_{ds}$), $\mu_{s/d}$ is the chemical potential at source/drain, and $f(E-\mu_{s/d})$ is the Fermi-Dirac distribution function at source/drain.
%

%%%%%%%%%%%%%%%%%%%%%%%%%-------- References -------%%%%%%%%%%%%%%%%%%%%%%%%%%%%%%%%%%%%%

%\newpage
\bibliographystyle{IEEEtranDOI}
\bibliography{References}

\end{document}